\documentclass[11pt,twoside]{article}

\usepackage{asp2006}
\usepackage{epsf}
\usepackage{lscape}
\usepackage{natbib}
\usepackage{graphicx}
\bibpunct{(}{)}{;}{a}{}{,}

\begin{document}
\title{Metal-poor Stars}   
\author{Anna Frebel}   
\affil{McDonald Observatory, University of Texas, Austin TX 78712-0259}    

\begin{abstract} 
The abundance patterns of metal-poor stars provide us a wealth of
chemical information about various stages of the chemical evolution of
the Galaxy.  In particular, these stars allow us to study the
formation and evolution of the elements and the involved
nucleosynthesis processes. This knowledge is invaluable for our
understanding of the cosmic chemical evolution and the onset of star-
and galaxy formation. Metal-poor stars are the local equivalent of the
high-redshift Universe, and offer crucial observational constraints on
the nature of the first stars.  This review presents the history of
the first discoveries of metal-poor stars that laid the foundation to
this field. Observed abundance trends at the lowest metallicities are
described, as well as particular classes of metal-poor stars such as
r-process and C-rich stars. Scenarios on the origins of the abundances
of metal-poor stars and the application of large samples of metal-poor
stars to cosmological questions are discussed.

\end{abstract}

\section{Introduction}  

The first stars that formed from the pristine gas left after the Big
Bang were very massive \citep{bromm02}. After a very short life time
these co-called Population\,III stars exploded as supernovae, which
then provided the first metals to the interstellar medium. All
subsequent generations of stars formed from chemically enriched
material. Metal-poor stars that are observable today are
Population\,II objects and belong to the stellar generations that
formed from non-zero metallicity gas. In their atmospheres these
objects preserve information about the chemical composition of their
birth cloud. They thus provide archaeological evidence of the earliest
times of the Universe. In particular, the chemical abundance patterns
provide detailed information about the formation and evolution of the
elements and the involved nucleosynthesis processes. This knowledge is
invaluable for our understanding of the cosmic chemical evolution and
the onset of star- and galaxy formation. Metal-poor stars are the
local equivalent of the high-redshift Universe. Hence, they also
provide us with observational constraints on the nature of the first
stars and supernovae. Such knowledge is invaluable for various
theoretical works on the early Universe.

Focusing on long-lived low-mass ($\sim0.8\,M_{\odot}$) main-sequence
and giant metal-poor stars, we are observing stellar chemical
abundances that reflect the composition of the interstellar medium during their star
formation processes.  Main-sequence stars only have a shallow
convection zone that preserves the stars' birth composition over
billions of years. Stars on the red giant branch have deeper
convection zones that lead to a successive mixing of the surface with
nuclear burning products from the stellar interior. In the lesser
evolved giants the surface composition has not yet been significantly
altered by any such mixing processes. The main indicator used to
determine stellar metallicity is the iron abundance, [Fe/H], which is
defined as \mbox{[A/B]}$ = \log_{10}(N_{\rm A}/N_{\rm B})_\star -
\log_{10}(N_{\rm A}/N_{\rm B})_\odot$ for the number N of atoms of
elements A and B, and $\odot$ refers to the Sun. With few exceptions,
[Fe/H] traces the overall metallicity of the objects fairly well.

\begin{figure}[!t]
\begin{center}
\includegraphics[width=9.7cm, clip=,bb=63 334 530 630]{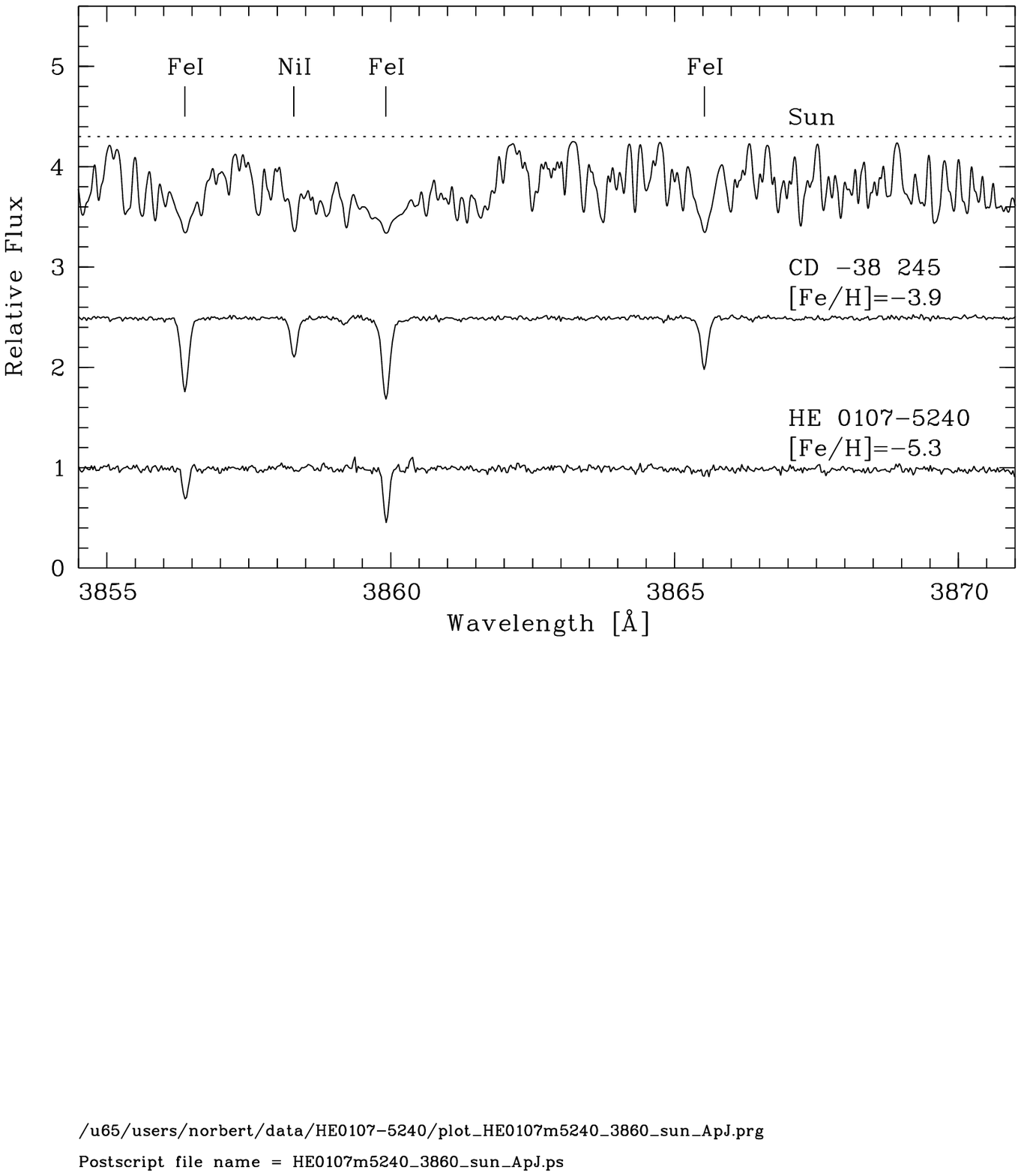}      
  \caption{\label{spec_comp_plot} Spectral comparison of the Sun with
  the metal-poor stars CD~$-38^{\circ}$ 245 and
  HE~0107$-$5240.  Figure taken from \citet{HE0107_ApJ}.}
\end{center}
\end{figure}

To illustrate the difference between younger metal-rich and older
metal-poor stars, Figure~\ref{spec_comp_plot} shows spectra of the Sun
and two of the most metal-poor stars,
CD~$-38^{\circ}$ 245 \citep{cd38} and HE~0107$-$5240
\citep{HE0107_Nature}. The number of the atomic absorption lines
detectable in the spectra decreases with increasing
metal-deficiency. In HE~0107$-$5240, only the intrinsically strongest
metal lines are left to observe. Compared with the metal-poor stars
presented in the figure, in a spectrum of a similarly unevolved
Population\,III object, no metal features would be detectable since it
contains no elements other than H, He and Li at its surface.

\begin{figure}[!t]
\begin{center}
\includegraphics[width=7.cm,clip=, bb=43 334 380 580]{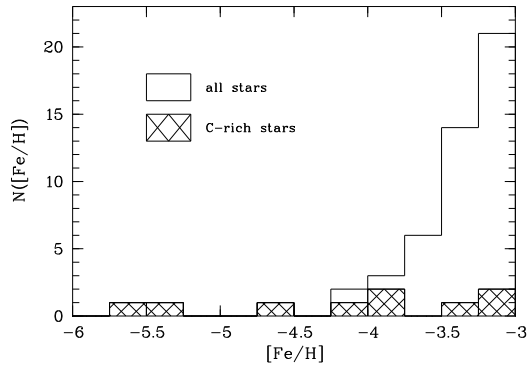}      
  \caption{\label{mdf} Low-metallicity tail of the observed halo
  metallicity distribution function. Only few objects are known, and
  many of the most iron-poor stars appear to be enriched with
  unusually large amounts of carbon. Figure adapted from \citet{he0557}.}
\end{center}
\end{figure}

Large numbers of metal-poor Galactic stars found in objective-prism
surveys in both hemispheres have provided enormous insight into the
formation and evolution of our Galaxy (e.g., \citealt{ARAA}). However,
there is only a very small number of stars known with metallicities
below $\mbox{[Fe/H]}<-3.5$. The number of known metal-poor stars
decreases significantly with decreasing metallicity as illustrated in
Figure~\ref{mdf}. Objects at the lowest metallicities are extremely
rare but of utmost importance for a full understanding of the early
Universe. Abundance trends are poorly defined in this metallicity
range, and the details of the metallicity distribution function (MDF)
at the lowest metallicity tail remains unclear. A major recent
achievement in the field was the push down to a new, significantly
lower limit of metallicity $\mbox{[Fe/H]}$ measured in a stellar
object: From a longstanding $\mbox{[Fe/H]}=-4.0$ (CD $-38^{\circ}$
245; \citealt{cd38}) down to $\mbox{[Fe/H]}=-5.3$ (HE~0107$-$5240;
\citealt{HE0107_Nature}), and very recently, down to
$\mbox{[Fe/H]}=-5.4$ (HE~1327$-$2326;
\citealt{HE1327_Nature}). Overall, only three stars are known with
iron abundances of $\mbox{[Fe/H]}<-4.0$. The recently discovered star
HE~0557$-$4840 \citep{he0557} with $\mbox{[Fe/H]}<-4.8$ bridges the
gap between $\mbox{[Fe/H]}=-4.0$ and the aforementioned two objects
with $\mbox{[Fe/H]}<-5.0$. These extremely iron-deficient objects have
opened a new, unique observational window of the time very shortly
after the Big Bang. They provide key insight into the very beginning
of the Galactic chemical evolution.

This review discusses the history of the first discoveries of
metal-poor stars that laid the foundation to this field in
\S~2, and observed abundance trends at the lowest
metallicities in \S~3. \S~4, \S~5,
and \S~6 describe particular classes of metal-poor
stars. The origins of metal-poor stars are reviewed in
\S~7 and the application of these stars to cosmological
questions in \S~8.

\section{First Discoveries of Metal-Poor Stars}\label{sec_disc}
It was long believed that all stars would have a similar chemical
composition to the Sun. In the late 1940's, however, some metal lines
observed in stars appeared to be unusually weak compared with the
Sun. It was first suspected that these stars could be
hydrogen-deficient or might have peculiar atmospheres, but
\citet{chamberlain} concluded that ``one possibly undesirable factor''
in their interpretation would be the ``prediction of abnormally small
amounts of Ca and Fe'' in these stars. They found about 1/20th of the
solar Ca and Fe values. Subsequent works on such metal-weak stars
during the following few decades confirmed that stars do indeed have
different metallicities that reflect different stages of the chemical
evolution undergone by the Galaxy. HD140283 was one object observed by
\citet{chamberlain}. Repeated studies in the past half century have
shown that this subgiant has 
$\mbox{[Fe/H]}\sim-2.5$ (e.g., \citealt{Norrisetal:2001}). HD140282 is
what one could call a ``classical'' metal-poor halo star that now
often serves as reference star in chemical abundance analyses.

In 1981, \citet{Bond1981} asked the question ``Where is
Population\,III?'' At the time, Population\,III stars were suggested
to be stars with $\mbox{[Fe/H]}<-3.0$ because no stars were known with
metallicities lower than 1/1000 of the solar iron abundance. Even at
$\mbox{[Fe/H]}\sim-3.0$, only extremely few objects were known
although more were predicted to exist by a simple model of chemical
evolution in the Galactic halo. \citet{Bond1981} reported on
unsuccessful searches for these Population\,III stars and concluded
that long-lived low-mass star could not easily form from
zero-metallicity gas, and hence were extremely rare, if not
altogether absent. Today, we know that star formation in
zero-metallicity gas indeed does not favor the creation of low-mass
stars due to insufficient cooling processes \citep{bromm02}. It has
also become apparent that the number of stars at the tail of the MDF
is extremely sparsely populated (Figure~\ref{mdf}). All those stars
are fainter, and hence further away, than those among which Bond was
searching ($B=10.5$ to 11.5).  As will be shown in this review, the
current lowest metallicity has well surpassed $\mbox{[Fe/H]}\sim-3.0$
although it is still unclear what the lowest observable metallicity of
halo stars may be.

A few years later, a star was serendipitously discovered, CD
$-38^{\circ}$ 245, with a record low Fe abundance of
$\mbox{[Fe/H]}=-4.5$ \citet{cd38}. It was even speculated that CD
$-38^{\circ}$ 245 might be a true Population\,III star, although
\citet{cd38} concluded an extreme Population\,II nature for the star.
The red giant was later reanalyzed and found to have
$\mbox{[Fe/H]}=-4.0$ \citep{Norrisetal:2001}. CD $-38^{\circ}$ 245
stayed the lowest metallicity star for almost 20 years.

Large objective-prism surveys were carried out since the early 1980ies
to systematically identify metal-poor stars. For a review on surveys
and search techniques, the reader is referred to
\citet{christlieb_review} and \citet{ARAA}.

\section{Abundance Trends at Low Metallicity}\label{sec_trends}
The quest to find more of the most metal-poor stars to investigate the
formation of the Galaxy lead to the first larger samples with
metallicities down to $\mbox{[Fe/H]}\sim-4.0$.  Abundance ratios
[X/Fe] as a function of [Fe/H] were extended to low metallicities for
the lighter elements ($Z<30$) and neutron-capture elements
($Z>56$). \citet{McWilliametal} studied 33 stars with
$-4.0<\mbox{[Fe/H]}<-2.0$.
The $\alpha$-elements Mg, Si, Ca, and Ti are enhanced by
$\sim0.4$\,dex with respect to Fe.  $\alpha$-elements are produced
through $\alpha$-capture during various burning stages of late stellar
evolution, before and during supernova explosions.
This enhancement is found down to the lowest metallicities, although
with few exceptions. Recently, some stars were discovered that are
$\alpha$-poor \citep{ivans_alphapoor}, whereas others are strongly
overabundant in Mg and Si \citep{aoki_mg}.  The iron-peak elements are
produced during supernova type\,II explosions and their yields depend on the
explosion energy. The abundance ratios of Cr/Fe and Mn/Fe become more
underabundant with decreasing [Fe/H] and below $\mbox{[Fe/H]=-2.5}$,
up to $\sim-0.7$ at $\mbox{[Fe/H]}\sim-3.5$ for Cr and $\sim-1.0$ for
Mn. Co becomes overabundant, up to $\sim0.8$. Sc/Fe and Ni/Fe have a
roughly solar ratio, even at $\mbox{[Fe/H]}\sim-4.0$. Compared with
these elements, the neutron-capture elements appear to behave
differently. Sr has an extremely large scatter ($\sim 2$\,dex),
indicating that different nucleosynthetic process may contribute to
its Galactic inventory. Ba has less but still significant scatter at
various metallicities. These abundance trends were confirmed by
\citet{ryan96} who used 22 stars in the same metallicity regime. These
authors furthermore proposed that the chemical enrichment of the
early interstellar medium was primarily due to the explosion energies
of the first supernovae that determine their yield distribution.

\begin{figure}[!t]
\begin{center}
 \includegraphics[width=11.5cm,clip=, bb=32 330 498 725]{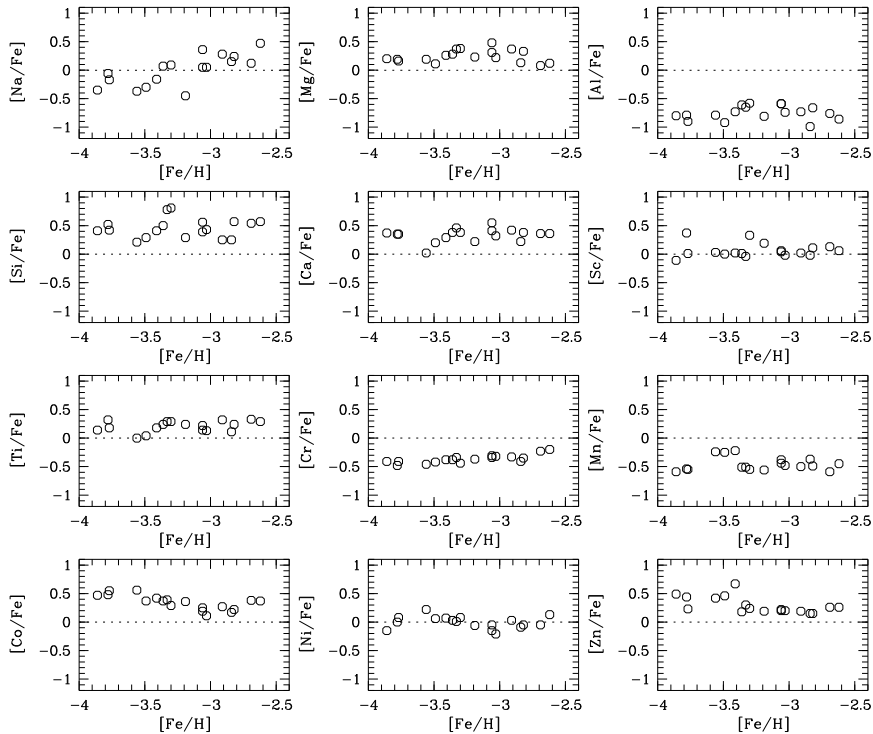}
   \caption{\label{cayrel_plot} Abundance trends at the lowest
   metallicities by means of the \citet{cayrel2004} stars.}
\end{center}
\end{figure}

Several stars at the lowest metallicities showed significant
deviations from the general trends. This level of abundance scatter at
the lowest metallicities led to the conclusion that the interstellar
medium was still inhomogeneous and not well mixed at
$\mbox{[Fe/H]}\sim-3.0$ (e.g., \citealt{argast}).

The claim for an inhomogeneous interstellar medium was diminished when
\citet{cayrel2004} presented a sample of 35 extremely metal-poor
stars. Figure~\ref{cayrel_plot} show the abundance trends for those
stars for 12 different elements.  These authors found very little
scatter in the abundance trends of elements with $Z<30$ among their
stars down to $\mbox{[Fe/H]}\sim-4.0$. This result suggested that the
interstellar medium was already well-mixed at these early time leading
to the formation of stars with almost identical abundance patterns.

\section{r-Process Stars}\label{sec_r}
\begin{figure}[!ht]
\begin{center}
 \includegraphics[width=8.cm,clip=,bb=84 14 490 390]{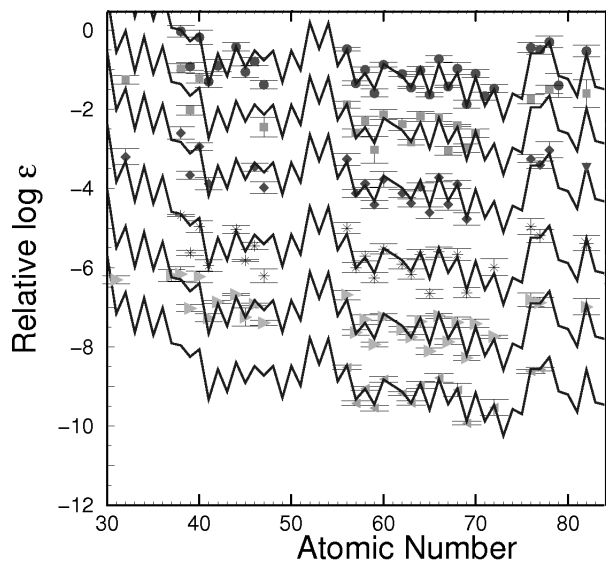}
   \caption{\label{r-process_plot} r-process abundance patterns of the
   most strongly-enhanced r-process metal-poor stars (various
   symbols). Overplotted are scaled solar r-process patterns. There is
   excellent agreement between the stellar data and the solar
   r-process pattern. All patterns are arbitrarily offset to allow a
   visual comparison. Figure kindly provided by J. J. Cowan.}
\end{center}
\end{figure}

All elements except H and He are created in stars during stellar
evolution and supernova explosions.  The so-called r-process stars
formed from material previously enriched with heavy neutron-capture
elements.  About $5\%$ of stars with $\mbox{[Fe/H]}<-2.5$ contain a
strong enhancement of neutron-capture elements associated with the
rapid (r-) nucleosynthesis process\citep{ARAA} that is responsible for
the production of the heaviest elements in the Universe. In those
stars we can observe the majority (i.e., $\sim70$ of 94) of elements
in the periodic table: the light, $\alpha$, iron-peak, and light and
heavy neutron-capture elements. So far, the nucleosynthesis site of
the r-process has not yet unambiguously been identified, but supernova
explosions are the most promising location. In 1996, the first
r-process star was discovered, CS~22892-052
\citep{Snedenetal:1996}. The heavy neutron-capture elements observed
in this star follow the scaled solar r-process pattern. The abundance
patterns of the heaviest elements with $56<Z<90$ of the few known
r-process stars indeed \textit{all} follow the scaled \textit{solar}
r-process pattern as can be seen in Figure~\ref{r-process_plot}. This
behavior suggests that the r-process is universal -- an important
empirical finding that could not be obtained from any laboratory on
earth.  However, there are deviations among the lighter
neutron-capture elements. Since it is not clear if the stellar
abundance patterns are produced by a single r-process only, an
additional new process might need to be invoked in order to explain
all neutron-capture abundances (e.g., \citealt{aoki05}).

Among the heaviest elements are the long-lived radioactive isotopes
$^{232}$Th (half-life $14$\,Gyr) or $^{238}$U ($4.5$\,Gyr). While Th
is often detectable in r-process stars, U poses a real challenge
because \textit{only one}, extremely weak line is available in the
optical spectrum. By comparing the abundances of the radioactive Th
and/or U with those of stable r-process nuclei, such as Eu, stellar
ages can be derived. Through individual age measurements, r-process
objects become vital probes for observational ``near-field'' cosmology
by providing an independent lower limit for the age of the Universe.
This fortuity also provides the opportunity of bringing together
astrophysics and nuclear physics because these objects act as ``cosmic
lab'' for both fields of study.

Most suitable for such age measurements are cool metal-poor giants
that exhibit strong overabundances of r-process
elements\footnote{Stars with [r/Fe] $>1.0$; r represents the average
abundance of elements associated with the r-process.}.  Since
CS~22892-052 is very C-rich, the U line is blended and not
detectable. Only the Th/Eu ratio could be employed, and an age of
$14$\,Gyr was derived \citep{sneden03}.  The U/Th chronometer was
first measured in the giant CS~31082-001 \citep{Cayreletal:2001},
yielding an age of $14$\,Gyr.
The recently discovered giant HE~1523$-$0901 ($\mbox{[Fe/H]}=-3.0$; \citealt{he1523})
has the largest measured overabundance of r-process elements. It is
only the \textit{third} star with a U detection at all, and is has the
most reliable measurement of all three. HE~1523$-$0901 is the first
star that has been dated with seven different ``cosmic clocks'',
i.e. abundance ratios such as U/Th, Th/Eu, U/Os. The average age
obtained is $13$\,Gyr, which is consistent with the WMAP
\citep{WMAP} age of the Universe of 13.7\,Gyr. These age measurements
also confirm the old age of similarly metal-poor stars.

Since Eu and Th are much easier to detect than U, the Th/Eu
chronometer has been used several times to derive stellar ages of
metal-poor stars. Table~\ref{ages} lists the ages derived from the
Th/Eu and other abundance ratios measured in the stars that are shown
in Figure~\ref{r-process_plot}.  Compared to Th/Eu, the Th/U ratio is
much more robust to uncertainties in the theoretically derived
production ratio due to the similar atomic masses of Th and U
\citep{wanajo2002}. Hence, stars displaying Th \textit{and} U are the
most desired stellar chronometers.

\begin{figure}[!t]
\begin{center}
\includegraphics[width=12.5cm,clip=, bb=31 112 530 315]{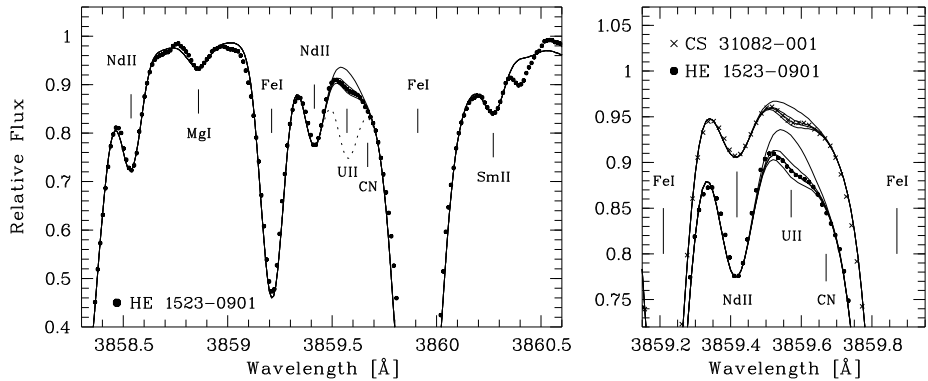}
  \caption{\label{U_region} Spectral region around the \ion{U}{II}
  line in HE~1523$-$0901 (\textit{filled dots}) and CS~31082-001
  (\textit{crosses}; right panel only). Overplotted are synthetic
  spectra with different U abundances. The dotted line in the left
  panel corresponds to a scaled solar r-process U abundance present in
  the star if no U were decayed. Figure taken from \citet{he1523}.}
\end{center}
\end{figure}

\begin{table}[!t]
\caption{\label{ages}Stellar ages derived from different abundance ratios}
\begin{center}
{\small
\begin{tabular}{lllll}
\tableline
\noalign{\smallskip}
Star  &Age (Gyr) &Abundance ratio &Ref. \\
\noalign{\smallskip}
\tableline
\noalign{\smallskip}

 HD15444     &$15.6\pm4$ & Th/Eu &\citet{cowan99}  \\ 
 CS31082-001 &$14.0\pm2$ &U/Th &\citet{Hilletal:2002}      \\  
 BD $+17^{\circ}$ 3248 &$13.8\pm4$ & average of several &\citet{cowan_U_02}\\
 CS22892-052 &$14.2\pm3$ & average of several &\citet{sneden03} \\  
 HE~1523$-$0901&$13.2\pm2$& average of several &\citet{he1523}\\
 HD22170     &$11.7\pm3$ & Th/Eu &\citet{ivans07}  \\

\noalign{\smallskip}
\tableline
\end{tabular}
}
\end{center}
\end{table}

\section{Carbon-Rich Stars \& s-Rich Stars}\label{sec_s}
It was first noted by \citet{1999rossicarbon} that a large fraction of
metal-poor stars has an overabundance of carbon with respect to iron
($\mbox{C/Fe}>1.0$).  This suggestion has been widely confirmed by now,
although different samples lead to a range of fractions (10\% to
25\%; e.g., \citep{cohen, lucatello2006}). At the lowest metalicities,
this fraction is even higher, and all three stars with
$\mbox{[Fe/H]}<-4.0$ are extremely C-rich. The fraction of C-rich
stars may also increase with increasing distance from the Galactic
plane \citep{frebel_bmps}.

Many C-rich stars also show an enhancement in neutron-capture
elements. These elements are produced in the interiors of
intermediate-mass asymptotic giant branch (AGB) stars through the slow
(s-) neutron-capture process. Such material is later dredged up to the
star's surface. Contrary to the r-process, the s-process is not
universal because two different sites seem to host s-process
nucleosynthesis.  The so-called ``weak'' component occurs in the
burning cores of the more massive stars, and preferentially produces
elements around $Z \sim 40$. The ``main'' component of the s-process
occurs in the helium shells of thermally pulsing lower mass AGB stars
and is believed to account for elements with $Z \ge 40$ (e.g.,
\citealt{arlandini1999}). The s-process leads to a different
characteristic abundance pattern than the r-process.  Indeed, the
s-process signature is observed in some metal-poor stars, and their
neutron-capture abundances follow the scaled solar s-process
pattern. The metal-poor objects observed today received the s-process
enriched material during a mass transfer event across a binary system
from their more massive companion that went through the AGB phase
(e.g., \citealt{1997norriscarbon,2001aokisprocess}).  The binary
nature of many s-process stars has been confirmed through monitoring
their radial velocity variations \citep{lucatello2005}.
\begin{figure}[!t]
\begin{center}
\includegraphics[width=7.5cm,clip=]{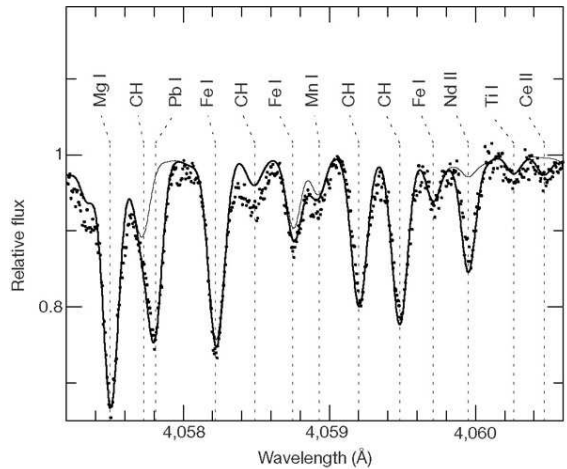}
  \caption{\label{Pb_region} Spectral region around the \ion{Pb}{I}
  line at $\sim$4058\,{\AA} in the metal-poor star HD196944
  (dots). Overplotted is the best-fit synthetic spectrum. Figure taken
  from \citet{2001vaneck}.}
\end{center}
\end{figure}
Some of the s-process metal-poor stars contain huge amounts of lead
(Pb) -- in fact they are more enhanced in lead than in any other
element heavier than iron \citep{2001vaneck}. Pb is the end product of
neutron-capture nucleosynthesis and AGB models predict an extended Pb,
and also Bi, production \citep{gallino1998}. More generally, the
observed patterns of many s-process metal-poor stars can be reproduced
with these AGB models, which is evidence for a solid theoretical
understanding of AGB nucleosynthesis.

Unlike the r-process, it is not clear whether the s-process already
operated in the early Universe, since it depends on the presence of
seed nuclei (i.e., from Fe-peak elements created in previous
generations of stars) in the host AGB star. However, there is recent
evidence from some metal-poor stars \citep{simmerer2004} that may
challenge the late occurrence of the s-process at higher
metallicities.

As usual, there are a exceptions. Some C-rich stars do not show any
s-process enrichment \citep{aoki_cempno, frebel_he1300}, and the
origin of these chemical signatures remains unclear. Other C-rich
stars show r-process enhancement \citep{Snedenetal:1996}, and the two
stars with the largest carbon overabundances with respect to iron are
the two with $\mbox{[Fe/H]}<-5.0$. The frequent finding of C-rich
stars in combination with a variety of abundance patterns points
towards the importance of C in the early Universe. The exact role of C
is still unclear but it likely played a crucial role in star formation
processes \citep{brommnature}.

\section{$\mbox{[Fe/H]}<-5.0$ Stars}\label{sec_5}
\begin{figure}[!t]
\begin{center}
\includegraphics[width=11.cm,clip=, bb=52 415 488 700]{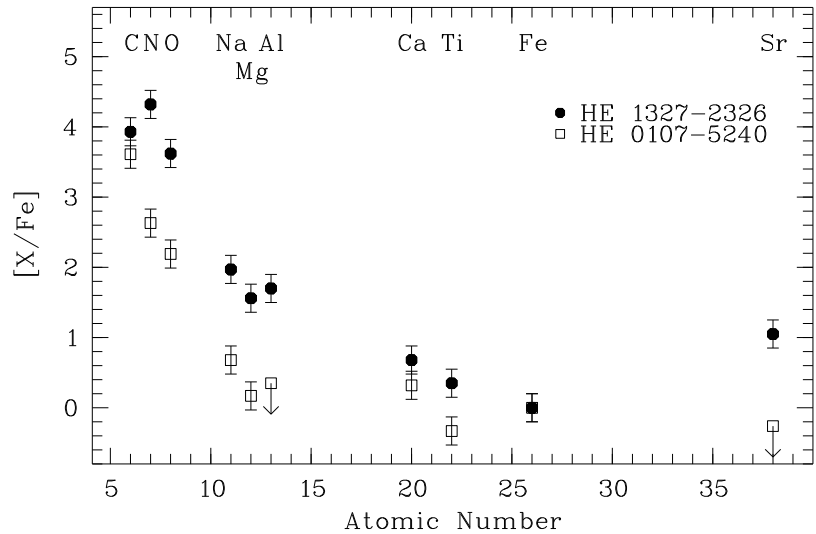}
  \caption{\label{abund_plot} Abundance patterns of HE~0107$-$5240 and
  HE~1327$-$2326. Abundances taken from
  \citet{HE0107_ApJ}, \citet{O_he0107}, \citet{HE1327_Nature}, and \citet{o_he1327}, where the
  LTE Fe abundance of HE~0107$-$5240 was corrected for non-LTE effects
  in the same way as for HE~1327$-$2326.}
\end{center}
\end{figure}

\begin{figure}[!t]
\begin{center}
\plottwo{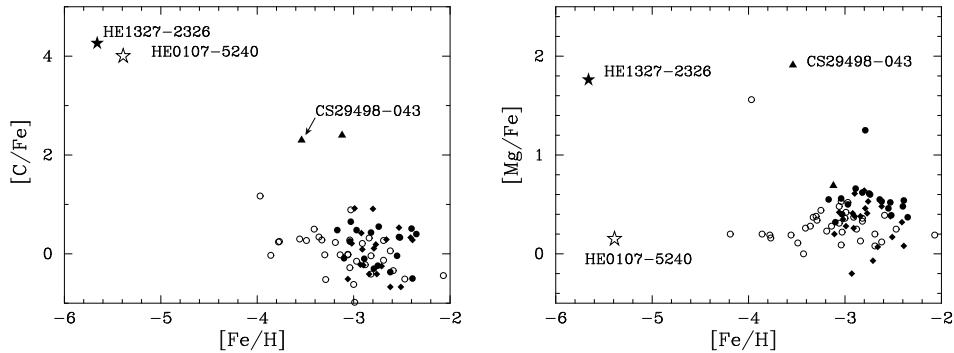}{f14.ps}
  \caption{\label{cfe_plot} C/Fe and Mg/Fe abundance ratios as function of
  [Fe/H]. See \citet{Aokihe1327} for full references.}
\end{center}
\end{figure}

In 2002, the first star with a new, record-low iron abundance was
found. The faint ($V=15.2$) red giant HE~0107$-$5240 has
$\mbox{[Fe/H]}=-5.3$ \citep{HE0107_Nature}\footnote{Applying the same
non-LTE correction of +0.2\,dex for Fe\,I abundances for
HE~0107$-$5240 and HE~1327$-$2326 leads to an abundance of
$\mbox{[Fe/H]}=-5.2$ for HE~0107$-$5240.}. In 2004, the bright
($V=13.5$) subgiant HE~1327$-$2326 was discovered
\citep{HE1327_Nature, Aokihe1327}. Both objects were discovered in the
Hamburg/ESO survey. HE~1327$-$2327 had an even lower iron abundance of
$\mbox{[Fe/H]}=-5.4$. This value corresponds to $\sim1/250,000$ of the
solar iron abundance. With its extremely large abundances of CNO
elements, HE~1327$-$2326 has a very similar abundance pattern compared
with HE~0107$-$5240 (see Figure~\ref{abund_plot}). No neutron-capture
element is found in HE~0107$-$5240, whereas, unexpectedly, Sr is
observed in HE~1327$-$2326. The Sr may originate from the
neutrino-induced $\nu p$-process operating in supernova explosions
\citep{froehlich}. Furthermore, in the relatively unevolved subgiant
HE~1327$-$2326 Li could not be detected, $\log\epsilon ({\rm
Li})<1.6$, where $\log\epsilon ({\rm A})$ = $\log_{10}(N_{\rm A}/N_{\rm
H})$ + 12. This is surprising, given that the primordial Li abundance
is often inferred from similarly unevolved metal-poor stars
\citep{ryan_postprim}. The upper limit found from HE~1327$-$2326,
however, strongly contradicts the WMAP value ($\log\epsilon ({\rm
Li})=2.6$) from the baryon-to-photon ratio \citep{WMAP}. This
indicates that the star must have formed from extremely Li-poor
material.  
Figure~\ref{cfe_plot} shows the C and Mg abundance for the
two stars in comparison with other, more metal-rich stars. Whereas
both $\mbox{[Fe/H]}<-5.0$ objects have similarly high C abundances
that are much higher than C/Fe ratios observed in other stars, there
exists a significant difference in the Mg abundances. HE~0107$-$5240
has an almost solar Mg/Fe ratio similar to the other stars. An
elevated Mg level is, however, observed not only in HE~1327$-$2326,
but also in a few stars at higher metallicities.

If HE~1327$-$2327 would have been a few hundred degrees hotter or a
few tens of a dex more metal-poor, no Fe lines would have been
detectable in the spectrum. Even now, the strongest Fe lines are
barely detectable and very high $S/N$ ratio ($>100$) was required for
the successful Fe measurement.  The hunt for the most metal-poor stars
with the lowest iron abundances is thus reaching a its technical limit
with regard to turn-off stars. Fortunately, in suitable giants, Fe
lines should be detectable at even lower Fe abundances than
$\mbox{[Fe/H]}<-5.5$. It remains to be seen, though, what the lowest
iron-abundance, or lowest overall metallicity for that matter, in a
stellar object might be.  Considering a simple accretion model,
\citep{iben1983} suggested such the observable lower limit to be
$\mbox{[Fe/H]}<-5.7$. Future observations may be able to reveal such a
lower limit.

HE~0107$-$5240 and HE~1327$-$2326 immediately became benchmark objects
to constrain various theoretical studies of the early Universe, such
as the formation of the first stars (e.g., \citealt{yoshida06}), the
chemical evolution of the early interstellar medium (e.g., \citealt{karlsson2005}) or
supernovae yields studies. More such stars with similarly low
metallicities are urgently needed. Very recently, a star was found at
$\mbox{[Fe/H]}<-4.75$ \citep{he0557}. It sits right in the previously
claimed ``metallicity gap'' between the stars at
$\mbox{[Fe/H]}\sim-4.0$ and the two with $\mbox{[Fe/H]}<-5.0$. These
three stars provide crucial information for the shape of the tail of
the metallicity distribution function.

\section{What is the Chemical Origin of the Most Metal-Poor Stars?}\label{sec_orig}
The two known hyper-metal-poor stars (with $\mbox{[Fe/H]}<-5.0$;
\citealt{HE0107_Nature, HE1327_Nature}) provide a new observational
window to study the onset of the chemical evolution of the Galaxy. The
\textit{highly individual} abundance patterns of these and other
metal-poor stars have been successfully reproduced by several
supernovae scenarios. HE~0107$-$5240 and HE~1327$-$2326 both appear to
be early, extreme Population\,II stars possibly displaying the
``fingerprint'' of only one Population\,III
supernova. \citet{UmedaNomotoNature} first reproduced the observed
abundance pattern of HE~0107$-$5240 by suggesting the star formed from
material enriched by a faint 25\,M$_{\odot}$ supernova that underwent
a mixing and fallback process.  To achieve a simultaneous enrichment
of a lot of C and only little Fe, large parts of the Fe-rich yield
fall back onto the newly created black hole.  Using yields from a
supernova with similar explosion energy and mass cut,
\citet{iwamoto_science} then explained the origin of HE~1327$-$2326.

\citet{meynet2005} explored the influence of stellar rotation on
elemental yields of very low-metallicity supernovae.  The stellar mass
loss yields of fast rotating massive Pop\,III stars quantitatively
reproduce the CNO abundances observed in HE~1327$-$2326 and other
metal-poor stars. \citet{limongi_he0107} were able to reproduce the
abundances of HE~0107$-$5240 through pollution of the birth cloud by
at least two supernovae. \citet{suda} proposed that the abundances of
HE~0107$-$5240 would originate from a mass transfer of CNO elements
from a potential companion, and from accretion of heavy elements from
the interstellar medium. However, neither for HE~0107$-$5240 nor
HE~1327$-$2326 radial velocity variations, that would indicate
binarity. Self-enrichment with CNO elements has also been ruled out
for HE~0107$-$524 \citep{picardi}, while this is not a concern for the
less evolved subgiant HE~1327$-$2326. 

\citet{tominaga07_b} model the averaged abundance pattern of four
non-carbon-enriched stars with $-4.2<\mbox{[Fe/H]}<-3.5$
\citep{cayrel2004} with the elemental yields of a massive, energetic
($\sim30-50$\,M$_{\odot}$) Population\,III hypernova. Abundance
patterns of stars with $\mbox{[Fe/H]}\sim-2.5$ can be reproduced with
integrated (over a Salpeter IMF) yields of normal Population\,III
supernovae.

\section{Stellar Archaeology}\label{sec_arch}
Numerical simulations show that the first stars in the Universe must
have been very massive ($\sim100 $\,M$_{\odot}$). On the other hand,
the observed metal-poor stars all have \textit{low masses}
($<1$\,M$_{\odot}$).  To constrain the first low-mass star formation,
\citet{brommnature} put forward their theory that fine-structure line
cooling by C and O nuclei, as provided by the Population\,III objects,
may be responsible for the necessary cooling of the interstellar medium to allow
low-mass stars to form. On the other hand, cooling by dust may play a
major role in the transition from Population\,III to Population\,II
star formation. Dust grains created during the first supernova explosions
may induce cooling and subsequent fragmentation processes that lead to
the formation of subsolar-mass stars (e.g., \citealt{schneider06}).

Such cooling theories can be tested with large numbers of
\textit{carbon and oxygen-poor} metal-poor stars. In order to form, a
critical metallicity of the interstellar medium is
required. Metal-poor stars with the lowest metallicities should thus
have values equal or higher than the critical metallicity. Suitable
metallicity indicators for the fine-structure line theory are C and O
abundances in the most metal-poor stars.  \citep{dtrans} collected
observational literature data of these two abundances in metal-poor
stars. 
\begin{figure}[!ht]
\begin{center}
\includegraphics[width=10.7cm,clip=, bb=72 450 398 685]{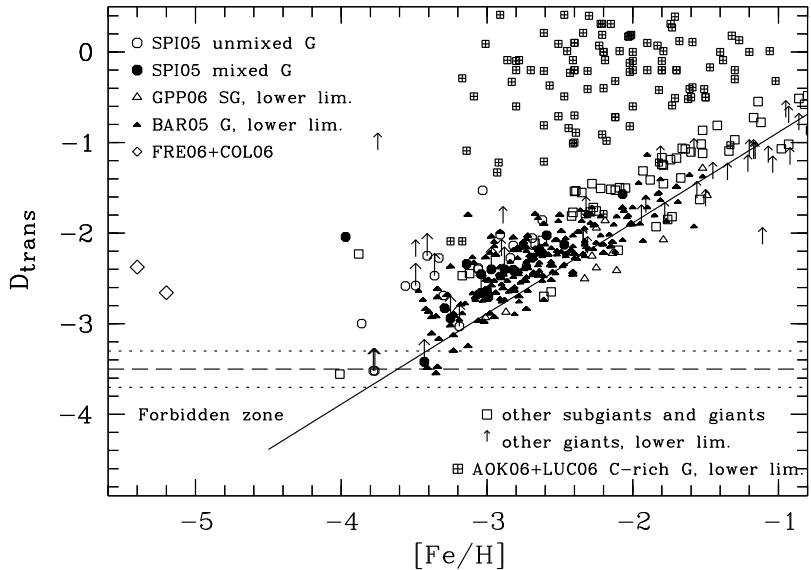} 
  \caption{\label{dtrans_plot} Comparison of metal-poor data with the theory that
  fine-structure lines of C and O dominate the transition from
  Population III to Population II in the early Universe.  C and O-poor
  extremely metal-poor stars are invaluable to test this formation
  scenario. Figure taken from \citet{dtrans}.}
\end{center}
\end{figure}
A critical metallicity limit of $D_{\rm trans}=-3.5$ was
developed, where $D_{\rm trans}$ parameterizes the observed C and/or O
abundances. So far, no star has abundances below the critical value,
as can be seen in Figure~\ref{dtrans_plot}.  It is predicted that any
future stars with $\mbox{[Fe/H]}<-4$ to be discovered should have high
C and/or O abundances. This agrees with the empirically frequent
C-richness amongst the most metal-poor stars. Since the critical
metallicity of the dust cooling theory is a few orders of magnitude
below the $D_{\rm trans}$ the current data is also consistent with
that theory, although less firm constraints can be derived at this
point.  Future observations of large numbers of metal-poor stars will
provide the necessary constraints that may lead to identify the
dominant cooling process responsible for low-mass star formation in
the early Universe.

\section{Outlook}\label{sec_summ}
More and more metal-poor stars are being discovered these days, thanks
to recent and ongoing large-scale surveys. The hunt for the most
metal-poor stars has certainly just begun with the discoveries of the
two stars with $\mbox{[Fe/H]}<-5.0$. Exciting times are ahead which
will hopefully lead to observations of new metal-poor stars from which
we can learn about the first stars, the onset of the Galactic chemical
enrichment and the involved nucleosynthesis processes. More stars at
the lowest metallicities are clearly desired in this quest and the
next-generations telescopes will certainly be of great interest to
this field. There are ongoing targeted searches for the most
metal-poor stars with existing 8-10\,m optical facilities. Large
high-resolution spectroscopy programs have been started with the
Hobby-Eberly-Telescope, \textit{Chemical Abundance of Stars in the
Halo} (CASH), and also with Subaru, Keck and Magellan.  More faint
stars (down to $V\sim17$) are currently being discovered with surveys
such as SDSS and SDSS-II (SEGUE). However, many stars are too faint to
be followed-up with high-resolution, high $S/N$ spectroscopy with
the current 8-10\,m telescopes. Future 25\,m class telescopes, such as the
Giant Magellan Telescope, will be required for discovering the most
metal-poor stars in the outskirts of the Galactic halo and nearby
dwarf galaxies.

\acknowledgements I would like to thank the McDonald Observatory Board
of Visitors and the meeting organizers for making possible such a
unique and inspiring meeting of young astronomers. I feel honored to
have been a part of it. Financial support through the W.~J.~McDonald
Fellowship of the McDonald Observatory is gratefully acknowledged.


\begin{thebibliography}{}
\expandafter\ifx\csname natexlab\endcsname\relax\def\natexlab#1{#1}\fi

\bibitem[{{Aoki} {et~al.}(2006){Aoki}, {Frebel}, {Christlieb}, {Norris},
  {Beers}, {Minezaki}, {Barklem}, {Honda}, {Takada-Hidai}, {Asplund}, {Ryan},
  {Tsangarides}, {Eriksson}, {Steinhauer}, {Deliyannis}, {Nomoto}, {Fujimoto},
  {Ando}, {Yoshii}, \& {Kajino}}]{Aokihe1327}
{Aoki}, W., {Frebel}, A., {Christlieb}, N., {Norris}, J.~E., {Beers}, T.~C.,
  {Minezaki}, T., {Barklem}, P.~S., {Honda}, S., {Takada-Hidai}, M., {Asplund},
  M., {Ryan}, S.~G., {Tsangarides}, S., {Eriksson}, K., {Steinhauer}, A.,
  {Deliyannis}, C.~P., {Nomoto}, K., {Fujimoto}, M.~Y., {Ando}, H., {Yoshii},
  Y., \& {Kajino}, T. 2006, ApJ, 639, 897

\bibitem[{{Aoki} {et~al.}(2005){Aoki}, {Honda}, {Beers}, {Kajino}, {Ando},
  {Norris}, {Ryan}, {Izumiura}, {Sadakane}, \& {Takada-Hidai}}]{aoki05}
{Aoki}, W., {Honda}, S., {Beers}, T.~C., {Kajino}, T., {Ando}, H., {Norris},
  J.~E., {Ryan}, S.~G., {Izumiura}, H., {Sadakane}, K., \& {Takada-Hidai}, M.
  2005, \apj, 632, 611

\bibitem[{{Aoki} {et~al.}(2002{\natexlab{a}}){Aoki}, {Norris}, {Ryan}, {Beers},
  \& {Ando}}]{aoki_mg}
{Aoki}, W., {Norris}, J.~E., {Ryan}, S.~G., {Beers}, T.~C., \& {Ando}, H.
  2002{\natexlab{a}}, \apjl, 576, L141

\bibitem[{{Aoki} {et~al.}(2002{\natexlab{b}}){Aoki}, {Norris}, {Ryan}, {Beers},
  \& {Ando}}]{aoki_cempno}
---. 2002{\natexlab{b}}, \apj, 567, 1166

\bibitem[{{Aoki} {et~al.}(2001){Aoki}, {Ryan}, {Norris}, {Beers}, {Ando},
  {Iwamoto}, {Kajino}, {Mathews}, \& {Fujimoto}}]{2001aokisprocess}
{Aoki}, W., {Ryan}, S.~G., {Norris}, J.~E., {Beers}, T.~C., {Ando}, H.,
  {Iwamoto}, N., {Kajino}, T., {Mathews}, G.~J., \& {Fujimoto}, M.~Y. 2001,
  \apj, 561, 346

\bibitem[{{Argast} {et~al.}(2000){Argast}, {Samland}, {Gerhard}, \&
  {Thielemann}}]{argast}
{Argast}, D., {Samland}, M., {Gerhard}, O.~E., \& {Thielemann}, F.-K. 2000,
  A\&A, 356, 873

\bibitem[{{Arlandini} {et~al.}(1999){Arlandini}, {K{\"a}ppeler}, {Wisshak},
  {Gallino}, {Lugaro}, {Busso}, \& {Straniero}}]{arlandini1999}
{Arlandini}, C., {K{\"a}ppeler}, F., {Wisshak}, K., {Gallino}, R., {Lugaro},
  M., {Busso}, M., \& {Straniero}, O. 1999, \apj, 525, 886

\bibitem[{{Beers} \& {Christlieb}(2005)}]{ARAA}
{Beers}, T.~C. \& {Christlieb}, N. 2005, ARAA, 43, 531

\bibitem[{{Bessell} {et~al.}(2004){Bessell}, {Christlieb}, \&
  {Gustafsson}}]{O_he0107}
{Bessell}, M.~S., {Christlieb}, N., \& {Gustafsson}, B. 2004, ApJ, 612, L61

\bibitem[{{Bessell} \& {Norris}(1984)}]{cd38}
{Bessell}, M.~S. \& {Norris}, J. 1984, ApJ, 285, 622

\bibitem[{{Bond}(1981)}]{Bond1981}
{Bond}, H. 1981, ApJ, 248, 606

\bibitem[{{Bromm} {et~al.}(2002){Bromm}, {Coppi}, \& {Larson}}]{bromm02}
{Bromm}, V., {Coppi}, P.~S., \& {Larson}, R.~B. 2002, ApJ, 564, 23

\bibitem[{{Bromm} \& {Loeb}(2003)}]{brommnature}
{Bromm}, V. \& {Loeb}, A. 2003, Nature, 425, 812

\bibitem[{{Cayrel} {et~al.}(2004){Cayrel}, {Depagne}, {Spite}, {Hill}, {Spite},
  {Fran{\c c}ois}, {Plez}, {Beers}, {Primas}, {Andersen}, {Barbuy},
  {Bonifacio}, {Molaro}, \& {Nordstr{\"o}m}}]{cayrel2004}
{Cayrel}, R., {Depagne}, E., {Spite}, M., {Hill}, V., {Spite}, F., {Fran{\c
  c}ois}, P., {Plez}, B., {Beers}, T., {Primas}, F., {Andersen}, J., {Barbuy},
  B., {Bonifacio}, P., {Molaro}, P., \& {Nordstr{\"o}m}, B. 2004, A\&A, 416,
  1117

\bibitem[{Cayrel {et~al.}(2001)Cayrel, Hill, Beers, Barbuy, Spite, Spite, Plez,
  Andersen, Bonifacio, Francois, Molaro, Nordstr{\"o}m, \&
  Primas}]{Cayreletal:2001}
Cayrel, R., Hill, V., Beers, T., Barbuy, B., Spite, M., Spite, F., Plez, B.,
  Andersen, J., Bonifacio, P., Francois, P., Molaro, P., Nordstr{\"o}m, B., \&
  Primas, F. 2001, Nature, 409, 691

\bibitem[{{Chamberlain} \& {Aller}(1951)}]{chamberlain}
{Chamberlain}, J.~W. \& {Aller}, L.~H. 1951, ApJ, 114, 52

\bibitem[{{Christlieb}(2006)}]{christlieb_review}
{Christlieb}, N. 2006, in Astronomical Society of the Pacific Conference
  Series, Vol. 353, Stellar Evolution at Low Metallicity: Mass Loss,
  Explosions, Cosmology, ed. H.~J.~G.~L.~M. {Lamers}, N.~{Langer}, T.~{Nugis},
  \& K.~{Annuk}, 271

\bibitem[{{Christlieb} {et~al.}(2002){Christlieb}, {Bessell}, {Beers},
  {Gustafsson}, {Korn}, {Barklem}, {Karlsson}, {Mizuno-Wiedner}, \&
  {Rossi}}]{HE0107_Nature}
{Christlieb}, N., {Bessell}, M.~S., {Beers}, T.~C., {Gustafsson}, B., {Korn},
  A., {Barklem}, P.~S., {Karlsson}, T., {Mizuno-Wiedner}, M., \& {Rossi}, S.
  2002, Nature, 419, 904

\bibitem[{{Christlieb} {et~al.}(2004){Christlieb}, {Gustafsson}, {Korn},
  {Barklem}, {Beers}, {Bessell}, {Karlsson}, \& {Mizuno-Wiedner}}]{HE0107_ApJ}
{Christlieb}, N., {Gustafsson}, B., {Korn}, A.~J., {Barklem}, P.~S., {Beers},
  T.~C., {Bessell}, M.~S., {Karlsson}, T., \& {Mizuno-Wiedner}, M. 2004, ApJ,
  603, 708

\bibitem[{{Cohen} {et~al.}(2005){Cohen}, {Shectman}, {Thompson}, {McWilliam},
  {Christlieb}, {Melendez}, {Zickgraf}, {Ram{\'{\i}}rez}, \& {Swenson}}]{cohen}
{Cohen}, J.~G., {Shectman}, S., {Thompson}, I., {McWilliam}, A., {Christlieb},
  N., {Melendez}, J., {Zickgraf}, F.-J., {Ram{\'{\i}}rez}, S., \& {Swenson}, A.
  2005, ApJ, 633, L109

\bibitem[{{Cowan} {et~al.}(1999){Cowan}, {Pfeiffer}, {Kratz}, {Thielemann},
  {Sneden}, {Burles}, {Tytler}, \& {Beers}}]{cowan99}
{Cowan}, J.~J., {Pfeiffer}, B., {Kratz}, K.-L., {Thielemann}, F.-K., {Sneden},
  C., {Burles}, S., {Tytler}, D., \& {Beers}, T.~C. 1999, ApJ, 521, 194

\bibitem[{{Cowan} {et~al.}(2002){Cowan}, {Sneden}, {Burles}, {Ivans}, {Beers},
  {Truran}, {Lawler}, {Primas}, {Fuller}, {Pfeiffer}, \& {Kratz}}]{cowan_U_02}
{Cowan}, J.~J., {Sneden}, C., {Burles}, S., {Ivans}, I.~I., {Beers}, T.~C.,
  {Truran}, J.~W., {Lawler}, J.~E., {Primas}, F., {Fuller}, G.~M., {Pfeiffer},
  B., \& {Kratz}, K.-L. 2002, \apj, 572, 861

\bibitem[{{Frebel} {et~al.}(2005){Frebel}, {Aoki}, {Christlieb}, {Ando},
  {Asplund}, {Barklem}, {Beers}, {Eriksson}, {Fechner}, {Fujimoto}, {Honda},
  {Kajino}, {Minezaki}, {Nomoto}, {Norris}, {Ryan}, {Takada-Hidai},
  {Tsangarides}, \& {Yoshii}}]{HE1327_Nature}
{Frebel}, A., {Aoki}, W., {Christlieb}, N., {Ando}, H., {Asplund}, M.,
  {Barklem}, P.~S., {Beers}, T.~C., {Eriksson}, K., {Fechner}, C., {Fujimoto},
  M.~Y., {Honda}, S., {Kajino}, T., {Minezaki}, T., {Nomoto}, K., {Norris},
  J.~E., {Ryan}, S.~G., {Takada-Hidai}, M., {Tsangarides}, S., \& {Yoshii}, Y.
  2005, Nature, 434, 871

\bibitem[{{Frebel} {et~al.}(2006{\natexlab{a}}){Frebel}, {Christlieb},
  {Norris}, {Aoki}, \& {Asplund}}]{o_he1327}
{Frebel}, A., {Christlieb}, N., {Norris}, J.~E., {Aoki}, W., \& {Asplund}, M.
  2006{\natexlab{a}}, ApJ, 638, L17

\bibitem[{{Frebel} {et~al.}(2006{\natexlab{b}}){Frebel}, {Christlieb},
  {Norris}, {Beers}, {Bessell}, {Rhee}, {Fechner}, {Marsteller}, {Rossi},
  {Thom}, {Wisotzki}, \& {Reimers}}]{frebel_bmps}
{Frebel}, A., {Christlieb}, N., {Norris}, J.~E., {Beers}, T.~C., {Bessell},
  M.~S., {Rhee}, J., {Fechner}, C., {Marsteller}, B., {Rossi}, S., {Thom}, C.,
  {Wisotzki}, L., \& {Reimers}, D. 2006{\natexlab{b}}, ApJ, 652, 1585

\bibitem[{{Frebel} {et~al.}(2007{\natexlab{a}}){Frebel}, {Christlieb},
  {Norris}, {Thom}, {Beers}, \& {Rhee}}]{he1523}
{Frebel}, A., {Christlieb}, N., {Norris}, J.~E., {Thom}, C., {Beers}, T.~C., \&
  {Rhee}, J. 2007{\natexlab{a}}, ApJ, 660, L117

\bibitem[{{Frebel} {et~al.}(2007{\natexlab{b}}){Frebel}, {Johnson}, \&
  {Bromm}}]{dtrans}
{Frebel}, A., {Johnson}, J.~L., \& {Bromm}, V. 2007{\natexlab{b}}, MNRAS, 380,
  L40

\bibitem[{{Frebel} {et~al.}(2007{\natexlab{c}}){Frebel}, {Norris}, {Aoki},
  {Honda}, {Bessell}, {Takada-Hidai}, {Beers}, \& {Christlieb}}]{frebel_he1300}
{Frebel}, A., {Norris}, J.~E., {Aoki}, W., {Honda}, S., {Bessell}, M.~S.,
  {Takada-Hidai}, M., {Beers}, T.~C., \& {Christlieb}, N. 2007{\natexlab{c}},
  ApJ, 658, 534

\bibitem[{{Fr{\"o}hlich} {et~al.}(2006){Fr{\"o}hlich},
  {Mart{\'{\i}}nez-Pinedo}, {Liebend{\"o}rfer}, {Thielemann}, {Bravo}, {Hix},
  {Langanke}, \& {Zinner}}]{froehlich}
{Fr{\"o}hlich}, C., {Mart{\'{\i}}nez-Pinedo}, G., {Liebend{\"o}rfer}, M.,
  {Thielemann}, F.-K., {Bravo}, E., {Hix}, W.~R., {Langanke}, K., \& {Zinner},
  N.~T. 2006, Physical Review Letters, 96, 142502

\bibitem[{{Gallino} {et~al.}(1998){Gallino}, {Arlandini}, {Busso}, {Lugaro},
  {Travaglio}, {Straniero}, {Chieffi}, \& {Limongi}}]{gallino1998}
{Gallino}, R., {Arlandini}, C., {Busso}, M., {Lugaro}, M., {Travaglio}, C.,
  {Straniero}, O., {Chieffi}, A., \& {Limongi}, M. 1998, ApJ, 497, 388

\bibitem[{Hill {et~al.}(2002)Hill, Plez, Cayrel, Nordstr{\" o}m, Andersen,
  Spite, Spite, Barbuy, Bonifacio, Depagne, Fran{\c c}ois, \&
  Primas}]{Hilletal:2002}
Hill, V., Plez, B., Cayrel, R., Nordstr{\" o}m, T. B.~B., Andersen, J., Spite,
  M., Spite, F., Barbuy, B., Bonifacio, P., Depagne, E., Fran{\c c}ois, P., \&
  Primas, F. 2002, A\&A, 387, 560

\bibitem[{{Iben}(1983)}]{iben1983}
{Iben}, I. 1983, Memorie della Societa Astronomica Italiana, 54, 321

\bibitem[{{Ivans} {et~al.}(2006){Ivans}, {Simmerer}, {Sneden}, {Lawler},
  {Cowan}, {Gallino}, \& {Bisterzo}}]{ivans07}
{Ivans}, I.~I., {Simmerer}, J., {Sneden}, C., {Lawler}, J.~E., {Cowan}, J.~J.,
  {Gallino}, R., \& {Bisterzo}, S. 2006, ApJ, 645, 613

\bibitem[{{Ivans} {et~al.}(2003){Ivans}, {Sneden}, {James}, {Preston},
  {Fulbright}, {H{\"o}flich}, {Carney}, \& {Wheeler}}]{ivans_alphapoor}
{Ivans}, I.~I., {Sneden}, C., {James}, C.~R., {Preston}, G.~W., {Fulbright},
  J.~P., {H{\"o}flich}, P.~A., {Carney}, B.~W., \& {Wheeler}, J.~C. 2003, \apj,
  592, 906

\bibitem[{{Iwamoto} {et~al.}(2005){Iwamoto}, {Umeda}, {Tominaga}, {Nomoto}, \&
  {Maeda}}]{iwamoto_science}
{Iwamoto}, N., {Umeda}, H., {Tominaga}, N., {Nomoto}, K., \& {Maeda}, K. 2005,
  Science, 309, 451

\bibitem[{{Karlsson} \& {Gustafsson}(2005)}]{karlsson2005}
{Karlsson}, T. \& {Gustafsson}, B. 2005, A\&A, 436, 879

\bibitem[{{Limongi} {et~al.}(2003){Limongi}, {Chieffi}, \&
  {Bonifacio}}]{limongi_he0107}
{Limongi}, M., {Chieffi}, A., \& {Bonifacio}, P. 2003, \apjl, 594, L123

\bibitem[{{Lucatello} {et~al.}(2006){Lucatello}, {Beers}, {Christlieb},
  {Barklem}, {Rossi}, {Marsteller}, {Sivarani}, \& {Lee}}]{lucatello2006}
{Lucatello}, S., {Beers}, T.~C., {Christlieb}, N., {Barklem}, P.~S., {Rossi},
  S., {Marsteller}, B., {Sivarani}, T., \& {Lee}, Y.~S. 2006, ApJ, 652, L37

\bibitem[{{Lucatello} {et~al.}(2005){Lucatello}, {Tsangarides}, {Beers},
  {Carretta}, {Gratton}, \& {Ryan}}]{lucatello2005}
{Lucatello}, S., {Tsangarides}, S., {Beers}, T.~C., {Carretta}, E., {Gratton},
  R.~G., \& {Ryan}, S.~G. 2005, \apj, 625, 825

\bibitem[{{McWilliam} {et~al.}(1995){McWilliam}, {Preston}, {Sneden}, \&
  {Searle}}]{McWilliametal}
{McWilliam}, A., {Preston}, G.~W., {Sneden}, C., \& {Searle}, L. 1995, AJ, 109,
  2757

\bibitem[{{Meynet} {et~al.}(2006){Meynet}, {Ekstr{\"o}m}, \&
  {Maeder}}]{meynet2005}
{Meynet}, G., {Ekstr{\"o}m}, S., \& {Maeder}, A. 2006, A\&A, 447, 623

\bibitem[{{Norris} {et~al.}(2007){Norris}, {Christlieb}, {Korn}, {Eriksson},
  {Bessell}, {Beers}, {Wisotzki}, \& {Reimers}}]{he0557}
{Norris}, J.~E., {Christlieb}, N., {Korn}, A.~J., {Eriksson}, K., {Bessell},
  M.~S., {Beers}, T.~C., {Wisotzki}, L., \& {Reimers}, D. 2007, ApJ, 670, 774

\bibitem[{{Norris} {et~al.}(1997){Norris}, {Ryan}, \&
  {Beers}}]{1997norriscarbon}
{Norris}, J.~E., {Ryan}, S.~G., \& {Beers}, T.~C. 1997, \apj, 488, 350

\bibitem[{Norris {et~al.}(2001)Norris, Ryan, \& Beers}]{Norrisetal:2001}
Norris, J.~E., Ryan, S.~G., \& Beers, T.~C. 2001, ApJ, 561, 1034

\bibitem[{{Picardi} {et~al.}(2004){Picardi}, {Chieffi}, {Limongi}, {Pisanti},
  {Miele}, {Mangano}, \& {Imbriani}}]{picardi}
{Picardi}, I., {Chieffi}, A., {Limongi}, M., {Pisanti}, O., {Miele}, G.,
  {Mangano}, G., \& {Imbriani}, G. 2004, ApJ, 609, 1035

\bibitem[{{Rossi} {et~al.}(1999){Rossi}, {Beers}, \&
  {Sneden}}]{1999rossicarbon}
{Rossi}, S., {Beers}, T.~C., \& {Sneden}, C. 1999, in ASP Conf. Ser. 165: The
  Third Stromlo Symposium: The Galactic Halo, 264

\bibitem[{{Ryan} {et~al.}(1996){Ryan}, {Norris}, \& {Beers}}]{ryan96}
{Ryan}, S.~G., {Norris}, J.~E., \& {Beers}, T.~C. 1996, \apj, 471, 254

\bibitem[{{Ryan} {et~al.}(1999){Ryan}, {Norris}, \& {Beers}}]{ryan_postprim}
---. 1999, ApJ, 523, 654

\bibitem[{{Schneider} {et~al.}(2006){Schneider}, {Omukai}, {Inoue}, \&
  {Ferrara}}]{schneider06}
{Schneider}, R., {Omukai}, K., {Inoue}, A.~K., \& {Ferrara}, A. 2006, \mnras,
  369, 1437

\bibitem[{{Simmerer} {et~al.}(2004){Simmerer}, {Sneden}, {Cowan}, {Collier},
  {Woolf}, \& {Lawler}}]{simmerer2004}
{Simmerer}, J., {Sneden}, C., {Cowan}, J.~J., {Collier}, J., {Woolf}, V.~M., \&
  {Lawler}, J.~E. 2004, \apj, 617, 1091

\bibitem[{{Sneden} {et~al.}(2003){Sneden}, {Cowan}, {Lawler}, {Ivans},
  {Burles}, {Beers}, {Primas}, {Hill}, {Truran}, {Fuller}, {Pfeiffer}, \&
  {Kratz}}]{sneden03}
{Sneden}, C., {Cowan}, J.~J., {Lawler}, J.~E., {Ivans}, I.~I., {Burles}, S.,
  {Beers}, T.~C., {Primas}, F., {Hill}, V., {Truran}, J.~W., {Fuller}, G.~M.,
  {Pfeiffer}, B., \& {Kratz}, K.-L. 2003, \apj, 591, 936

\bibitem[{Sneden {et~al.}(1996)Sneden, McWilliam, Preston, Cowan, Burris, \&
  Amorsky}]{Snedenetal:1996}
Sneden, C., McWilliam, A., Preston, G.~W., Cowan, J.~J., Burris, D.~L., \&
  Amorsky, B.~J. 1996, ApJ, 467, 819

\bibitem[{{Spergel} {et~al.}(2007){Spergel}, {Bean}, {Dor{\'e}}, {Nolta},
  {Bennett}, {Dunkley}, {Hinshaw}, {Jarosik}, {Komatsu}, {Page}, {Peiris},
  {Verde}, {Halpern}, {Hill}, {Kogut}, {Limon}, {Meyer}, {Odegard}, {Tucker},
  {Weiland}, {Wollack}, \& {Wright}}]{WMAP}
{Spergel}, D.~N., {Bean}, R., {Dor{\'e}}, O., {Nolta}, M.~R., {Bennett}, C.~L.,
  {Dunkley}, J., {Hinshaw}, G., {Jarosik}, N., {Komatsu}, E., {Page}, L.,
  {Peiris}, H.~V., {Verde}, L., {Halpern}, M., {Hill}, R.~S., {Kogut}, A.,
  {Limon}, M., {Meyer}, S.~S., {Odegard}, N., {Tucker}, G.~S., {Weiland},
  J.~L., {Wollack}, E., \& {Wright}, E.~L. 2007, ApJS, 170, 377

\bibitem[{{Suda} {et~al.}(2004){Suda}, {Aikawa}, {Machida}, {Fujimoto}, \&
  {Iben}}]{suda}
{Suda}, T., {Aikawa}, M., {Machida}, M.~N., {Fujimoto}, M.~Y., \& {Iben}, I.~J.
  2004, ApJ, 611, 476

\bibitem[{{Tominaga} {et~al.}(2007){Tominaga}, {Umeda}, \&
  {Nomoto}}]{tominaga07_b}
{Tominaga}, N., {Umeda}, H., \& {Nomoto}, K. 2007, ApJ, 660, 516

\bibitem[{Umeda \& Nomoto(2003)}]{UmedaNomotoNature}
Umeda, H. \& Nomoto, K. 2003, Nature, 422, 871

\bibitem[{{Van Eck} {et~al.}(2001){Van Eck}, {Goriely}, {Jorissen}, \&
  {Plez}}]{2001vaneck}
{Van Eck}, S., {Goriely}, S., {Jorissen}, A., \& {Plez}, B. 2001, \nat, 412,
  793

\bibitem[{{Wanajo} {et~al.}(2002){Wanajo}, {Itoh}, {Ishimaru}, {Nozawa}, \&
  {Beers}}]{wanajo2002}
{Wanajo}, S., {Itoh}, N., {Ishimaru}, Y., {Nozawa}, S., \& {Beers}, T.~C. 2002,
  ApJ, 577, 853

\bibitem[{{Yoshida} {et~al.}(2006){Yoshida}, {Omukai}, {Hernquist}, \&
  {Abel}}]{yoshida06}
{Yoshida}, N., {Omukai}, K., {Hernquist}, L., \& {Abel}, T. 2006, ApJ, 652, 6

\end{thebibliography}

\end{document}